# Pursuit of thermoelectric properties in a novel Half Heusler compound: HfPtPb

Kulwinder Kaur[1], D. P. Rai[2], R. K. Thapa[3] and Sunita Srivastava[1*].

*[1]Department of Physics, Panjab University, Chandigarh, India*

*[2]Department of Physics, Pachhunga University College, Aizawl, India,*

*[3]Department of Physics, Mizoram University, Aizawl, India*

***Corresponding Author**:sunita@pu.ac.in*

**Abstract:** We explore the structural, electronic, mechanical and thermoelectric properties of a new half Heusler compound, HfPtPb which is all metallic heavy element and has been recently been proposed to be stable [Nature Chem 7 (2015) 308]. In the present work, we employ density functional theory and semi-classical Boltzmann transport equations with constant relaxation time approximation. The mechanical properties such as Shear modulus, Young's modulus, elastic constants, Poisson's ratio, and shear anisotropy factor are investigated. The elastic and phonon properties reveal that this compound is mechanically and dynamically stable. Pugh's and Frantsevich's ratio demonstrates the ductile behavior and Shear anisotropic factor reflects the anisotropic nature ofHfPtPb. The calculation of band structure predicts that this compound is semiconductor in nature with band gap 0.86 eV. The thermoelectric transport parameters such as Seebeck coefficient, electrical conductivity, and electronic thermal conductivity and lattice thermal conductivity have been calculated as a function of temperature. The highest value of Seebeck coefficient is obtained for n-type doping at optimal carrier concentration ($10^{20}$ e/cm$^3$).We predict the maximum value of figure of merit (0.25) at 1000 K. Our investigation suggests that this material is n-type semiconductor.

## 1.1 Introduction:

In the present era of tremendous industrial growth and over utilization of natural resources, there is an ever-increasing demand of energy which necessitates and compels us not only to look for alternative energy resources but also to take adequate steps to reduce the energy consumption. In fact, a large part of energy consumption is lost in the form of heat and the efforts are being made to exploit those energy materials which act smartly in response to the external fields like light, temperature and pressure. Types of materials which work on the basis of temperature gradient are thermoelectric (TE) materials. These absorb waste heat and convert it into electrical energy or vice versa. In last few decades, several materials have been studied experimentally as well as theoretically and some have been found to be highly efficient such as $Bi_2Te_3$, $Sb_2Te_3$, SnSe, PbTe, PbS [1-4], etc. However, under conditions of high temperatures, oxidation, and high toxicity, their poor chemical and physical stability limits their technological applications [5-6]. Over the last few decades the research interest has been focused towards the half-Heusler (HH) compounds not only due to their mechanically and thermally robust properties [7-10] but also because of their environment friendly constituents.

Among all HH compounds, those with total 18 valence electron counts are of particular interest as they exhibit semiconducting behavior in accordance with Slater Pauling rule [11]. An analysis of electronic band structure calculation suggests that the HH alloys with narrow band gaps are efficient TE materials [12-13]. The efficiency of a thermoelectric material is measured in terms of dimensionless figure of merit $ZT = S^2\sigma T/\kappa$, where S is the Seebeck coefficient, $\sigma$ is the electrical conductivity, and $\kappa$ is the total thermal conductivity constituting sum of contributions of electron ($k_e$) and lattice ($k_l$) thermal conductivities [14]. Efforts to optimize ZT can be carried out by two methods (i) phonon-phonon scattering and (ii) electronic-band engineering. Both Seebeck coefficient and phonon scattering could be enhanced by doping heavy elements which could modify the band energy near Fermi level. The intermetallic HH semiconductor such as NiTiSn and (Ti,Zr)CoSb have emerged as successful thermoelectric (TE) materials with large Seebeck coefficients, large thermopower and moderate electrical conductivity [15-16]. However, the major concern is their relatively high value of $k_l$ which is of the order of 10-15 (W/mK) which affects the efficiency for thermoelectric conversion [17-19].

Hohl et al. [20] reported a reduction of $k_l$ from 10 to 6 (W/mK) in ZrNiSn with Hf doping at Zr-site. This decrease in value of lattice thermal conductivity is due to the mass disorder at Zr-lattice that creates phonon scattering. It was also reported that the substitution of heavier atoms like Sb on Sn site in HfNiSn is an effective measure [21] to reduce the lattice thermal conductivity to 5–6 W/mK at room temperature, a heavily doped materials $Fe_{0.99}Co_{0.01}V_{0.6}Nb_{0.4}Sb$ gives 4.6 (W/mK) [22]. Similarly Du et al. [23] have also found reduced $k_l$ of 2.5 W/mK in $Hf_{0.25}Zr_{0.25}NiSn_{0.998}Sb_{0.002}$ from their experimental studies and ZT value close to ~1 at 718 K. Other than doping, a look for pure material with high phonon scattering ability is another choice to reduce the value of $k_l$. Recently, a suppressed $k_l$ of 5.0-3.5 W/mK (at 300K), 1.72 W/mK (at 300K) and 5.3 W/mK (at 700 K) were reported in undoped (Ho,Er,Dy)PdSb [24], LaPtSb[25] and NiTiSn [26], respectively. Some of the examples of recently developed HH materials with high ZT values are 2.2 for LaPtSb [27], 1.1 for ZrNiSn at 1100 K [28], 1.5 at 1200 K for FeNbSb [29], 1.38-3.54 for NaMgX (X=Pb, Sb, As) [30] etc. The aforementioned high thermoelectric performances of undoped HH alloy provide incentives to carry out more research on analogous HH compounds. One such compound can be HfPtPb whose detailed investigation of structural, electronic, thermodynamic and thermoelectric properties have not been carried out till date to best of our knowledge. The purpose of present study is to comprehensively analyze of the electronic and thermoelectric properties of HfPtPb.

The paper has been organized as follows: In the following section, calculation details have been discussed. In subsequent sections, calculations for electronic and thermoelectric properties have been presented. Finally, we summarize the results and present its analysis.

## 1.2 Calculation Details:

The calculations were carried out by using the pseudo potential plane wave method implemented in Quantum Espresso simulation package [31] simulation package. The electron exchange-correlation potential was treated within generalized gradient approximations (GGA) in the form of Perdew-Burke-Ernzerhof scheme [32]. The first Brillouin-zone integration was approximated by 1000 optimized k-points that generate 10x10x10 Monkhorst-Pack meshes [33]. The energy convergence criterion for self-consistent calculation is $10^{-4}$ Ry. The ground state optimized lattice constant was calculated by relaxing the unit cell volume based on Murnaghan's equation of state [34]. The lattice dynamical properties have been calculated in term of phonon frequencies as a second-order derivative of the total energy with respect to atomic displacements within the frame work of DFPT [35].

BoltzTraP code [36] is based on Boltzmann theory and calculates various band structure dependent quantities such as electrical conductivity (σ) and electronic thermal conductivity ($k_e$) within constant time approximation and rigid band approximation [RBA] [37, 38]. BoltzTraP code can analytically represent these band energies with a smoothed Fourier interpolation and thereafter we can obtain the necessary derivatives such as electron velocities for transport properties. The electrical conductivity (σ) and Seebeck coefficient (S) as function of group velocity ($v_\alpha$) are expressed as

$$\sigma_{\alpha\beta}(\varepsilon) = \frac{e^2}{N} \sum_{i,k} \tau v_\alpha(i,k) \frac{\delta(\varepsilon - \varepsilon_{i,k})}{d\varepsilon} \qquad (1)$$

$$S_{ij} = E_i (\overline{V_j} T)^{-1} = (\sigma^{-1}) \alpha_i v_{\alpha j} \qquad (2)$$

where N, τ, ε, α and β, $v_\alpha(i,k)$ are the number of k points sampled, relaxation time, band energy, tensor indices and a component of the group velocities respectively. Further group velocity could be written as

$$v_\alpha(i,k) = \frac{1}{\hbar} \frac{\partial \varepsilon_{i,k}}{\partial k_\alpha} \qquad (3)$$

The transport coefficients are a function of temperature and chemical potential and can be calculated by integrating the transport distribution [37]:

$$v_{\alpha\beta}(T; \mu) = \frac{1}{eT \Omega \sigma_{\alpha\beta}(T,\mu)} \int \sigma_{\alpha\beta}(\varepsilon)(\varepsilon - \mu) [-\frac{\partial f_\mu(T;\varepsilon)}{\partial \varepsilon}] d\varepsilon \qquad (4)$$

$$\kappa_{\alpha\beta}^0(T; \mu) = \frac{1}{e^2 T \Omega} \int \sigma_{\alpha\beta}(\varepsilon)(\varepsilon - \mu)^2 [-\frac{\partial f_\mu(T;\varepsilon)}{\partial \varepsilon}] d\varepsilon \qquad (5)$$

where e, $k^0$, Ω, μ and $f_\mu$ are the electronic charge, the electronic part of thermal conductivity, volume of unit cell, chemical potential, and Fermi –Dirac distribution function, respectively. The electrical conductivity (σ) is expressed in terms of the ratio of σ/τ. To calculate the electrical conductivity of the system, we have used the relaxation time τ=10⁻¹⁴ sec.

We have also used ShengBTE [39] code, based on phonon Boltzmann transport equation (pBTE) for the calculation of lattice thermal conductivity for further confirmation. This code includes second-order (harmonic) and third-order (anharmonic) interatomic force constants [40]. The methodology used in this work has been discussed elsewhere [41,42].

## 1.3 Results and Discussion:

## 1.3.1 Electronic properties

We start our discussion with the crystal structure of HfPtPb. Pristine HfPtPb is a face centered cubic structure with space group F-43m. The Wyckoff atomic positions are 4a site (0, 0, 0) occupied by Hf, while the 4c (1/4, 1/4, 1/4) and 4b site (1/2, 1/2, 1/2) are occupied by Pt and Pb, respectively. The remaining 4d (3/4, 3/4, 3/4) position is empty. The calculated lattice constant is 6.495 Å in good agreement with the previous available data, 6.485 Å [43]. The calculated lattice constant, Bulk modulus, pressure derivative of bulk modulus and band gap of this material is summarized in Table1.

**Table 1:** Calculated lattice constant, Bulk Modulus, Pressure derivative and band gap of HfPtPb

| Properties | Our work | Others [43] |
|---|---|---|
| Lattice constant (Å) | 6.495 | 6.485 |
| Bulk modulus (GPa) | 113.1 | - |
| Pressure derivative of B(GPa) | 4.35 | - |
| Band gap, $E_g$ (eV) | 0.86 | 0.97 |

The magnetic, non-magnetic and semiconducting performance of a HH alloy can easily be predicted from the Slater-Pauling equation [11], expressed as $M_t = (Z_t - 18)\mu_B$ for hybridization of the orbitals. Here $M_t$ and $Z_t$ are the total magnetic moment and the total number of valence electrons in the unit cell, respectively. In our calculations, the electronic configuration of constituent atoms are Hf [$5d^2 6s^2$], Pt[$5d^9 6s^1$] and Pb[$6s^2 6p^2$] which give total of 18 valence electrons. Thus, we can presume that HfPtPb is a nonmagnetic semiconductor with magnetic moment 0.0 $\mu_B$. The calculated band structure and density of states (DOS) are depicted in Fig 1. As shown in Fig.1(a), the valence band maxima (VBM) and conduction band minima (CBM) lie one above the other on $\Gamma$ high symmetry point which indicates that HfPtPb is direct band gap semiconductor. The reported value of band gap is 0.82 eV which gives reasonable agreement with the value reported by R. Gautier et al. [43]. The total density of states (DOS) and projected density of states (PDOS) confirms that HfPtPb is a semiconductor in nature. The projected density of states shows that the majority of density of states (DOS) is due to the dispersed d-states of Hf and Pt near the Fermi energy, while Pb-p is negligible. The d-orbitals of Pt give maximum contribution in valence band and d-orbitals of Hf gives maximum contribution in conduction band. *s* and *p*- orbitals of Pb show minor contribution in valence band as well as conduction band. The present of sharp band in conduction band indicate that these materials have high power factor in n-type composition. From the above discussion, we can say that d-electrons of elements play a major role in thermoelectric properties.

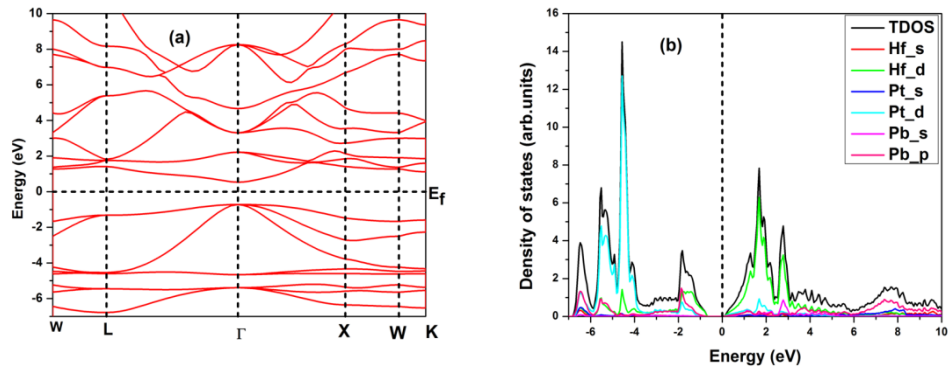

**Fig. 1**: (a) band structure and (b) density of states of HfPtPb

### 1.3.2 Mechanical properties:

The calculated mechanical properties like elastic constants, Shear modulus (G) and Bulk modulus (B) are listed in Table 2. The elastic constant ($C_{11}$) describe the stiffness against principal strains and $C_{44}$ shows the resistance against shear deformation. A fragile material has large value of G/B which is unfavorable for thermoelectric application. The mechanical properties of HfPtPb are checked by using Born-Huang stability criteria [44]:

$$C_{11}\text{-}C_{12}>0, \ C_{11}>0, \ C_{44}>0 \text{ and } (C_{11}+2C_{12})>0 \qquad (6)$$

The elastic constants in Table 2 satisfy the Born-Huang stability criteria. Therefore, according to this method HfPtPb is mechanically stable. The bulk modulus (B) and shear modulus (G) of HfPtPb are estimated with the help of Voigt-Reuss-Hill approximations (VRH) [45,46]. The value of bulk modulus (B) obtained with the help of VRH (114.2 GPa) is in good agreement with the value of bulk modulus obtained by EOS (113.1 GPa) which shows the good self consistency of our work. The value of Shear anisotropy (A), Poisson's ratio (v) and Young's modulus (Y) (Table 2) are calculated by using these equations:

$$A = \frac{2C_{44}}{C_{11} - C_{12}} \qquad (7)$$

$$\nu = \frac{(3B - 2G)}{2(3B + G)} \qquad (8)$$

$$Y = \frac{9BG}{3B + G} \qquad (9)$$

If the value of Shear anisotropy is equal to 1, then the material will be isotropic otherwise anisotropic behavior. The calculated value of Shear anisotropy factor [47] is 1.35 which is greater than 1 [48]. Therefore, this compound is purely anisotropic material. The obtained value of Poisson's ratio is 0.29, which concludes that interatomic forces of this compound are central forces [49]. The ratio of bulk and shear modulus is called Pugh's ratio [50], which gives the estimate about brittle and ductile nature of the materials. If Pugh's ratio > 1.75 [50], then the material will be ductile in nature, otherwise corresponds to brittle nature. Since the Pugh's ratio is 2.14, this means that this compound is ductile in nature. The inverse of Pugh's ratio is called Frantsevich's ratio (G/B). The calculated value of Frantsevich's ratio is 0.46 which is less than 1.06 [51] indicating that these materials have low resistance against shear deformation and hence are ductile in nature.

**Table 2. Thermodynamic and elastic properties of HfPtPb**

| Properties | Calculated values | Properties | Calculated values |
|---|---|---|---|
| $C_{11}$(GPa) | 185.63 | G(GPa) | 53.15 |
| $C_{12}$(GPa) | 97.07 | Shear anisotropy A | 1.35 |
| $C_{44}$(GPa) | 60.11 | Y(GPa) | 138.24 |
| $C_{11}$-$C_{12}$(GPa) | 88.56 | Poisson's Ratio ν | 0.29 |
| $C_{11}$+2$C_{12}$(GPa) | 379.77 | Pugh's ratio B/G | 2.14 |
| B(GPa) | 114.2 | Frantsevich's ratio G/B | 0.46 |

### 1.3.3 Phonon properties:

In order to investigate the structural stability of the system, a phonon dispersion curve has been plotted based on the harmonic force constants as obtained in Quantum Espresso package. The result of phonon band and phonon DOS has been presented in Fig.2. The high-symmetry phonon dispersion band occurs along the Γ-point [Fig. 2]. The presence of phonon dispersion in positive frequency range confirms the thermodynamic stability of our system [Fig.2]. The three atoms in the primitive cell of HfPtPb yield nine phonon modes (three acoustics and six optical modes) similar to HfNiPb [52]. The maximum contribution to the heat transfer comes from the acoustic modes due to their strong dispersion and large group velocities. The maximum amplitude of acoustic frequency is ~100 cm⁻¹, in good agreement with 120 cm⁻¹ of HfNiPb [52]. The mixing between the acoustic and the optical branches gives a considerable phonon-phonon scattering, which results in low thermal conductivity of the compound[53].

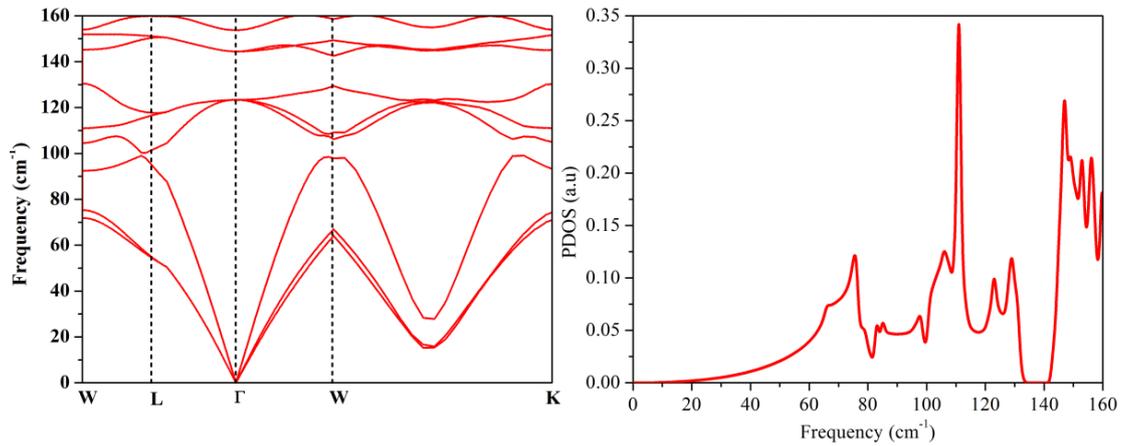

**Fig.2**: Phonon dispersion (band and DOS)

### 1.3.4 Thermoelectric properties

The variation of Seebeck coefficient of this material with temperature and concentration is depicted in Fig 3. The maximum value of Seebeck Coefficient is obtained for n-type doping and the optimal carrier concentration is $1.0 \times 10^{20}$ cm⁻³. The calculated Seebeck coefficient (S), electrical conductivity (σ), electronic thermal conductivity ($k_e$) and lattice thermal conductivity ($k_l$) as function of temperature at optimum value of $10^{20}$ cm⁻³ carrier concentrations are plotted in Fig 4. The sharp density of states near the band edge and the width of band gap are good sign for the thermoelectric properties of materials. In this material, the calculated band structure has a sharp conduction band and flat valence band which reveals that this material has good transport properties

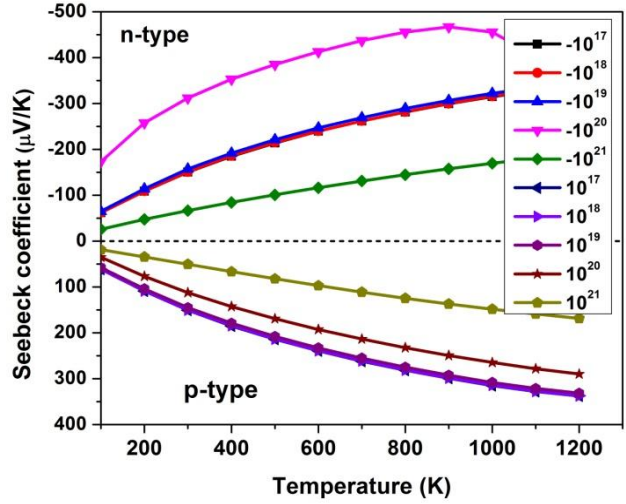

**Fig 3:** Variation of Seebeck coefficient of HfPtPb at different concentrations (cm$^{-3}$) and temperature

The ZT value above ~1 is the benchmark for the materials to exhibit good practical applications. In Fig. 4(a), it is observed that with increases in temperature, the value of Seebeck coefficient increases and at high temperature the value of Seebeck coefficient decreases because of presence of minority charge carriers. The negative value of Seebeck coefficient indicates that the charge carriers are electron. The material is n-type semiconductor. At room temperature, the calculated value of Seebeck coefficient is -312.09µV/K which attains peak value at 1000 K. Fig. 4(b) depicts the electrical conductivity (σ) to have a decreasing trend at low temperature which could be easily explained as follows. The electrical conductivity depends directly upon carrier concentration and mobility. However, as the mobility (µ) is inverse function of the temperature, the electrical conductivity is expected to decrease at low temperature. At 1000 K, the electrical conductivity begins to increase with temperature. The room temperature value of σ is 7.32x10$^{3}$(Ωm)$^{-1}$. Fig 4(c) shows that, the electronic thermal conductivity increases with increase in temperature. Fig 4(d) shows the variation of lattice thermal conductivity with temperature. The lattice thermal conductivity shows inverse temperature dependence. The room temperature value of $k_l$ is found to be 9.9 W/mK. Our calculated results of k$_l$ are also in qualitative agreement with the previously reported results of 5.0-3.5 W/mK for (Ho,Er,Dy)PdSb [24], 5.3 W/mK for NiTiSn [26], 6.7-10 W/mK for HfNiSn [54] and 8.0-7.6 W/mK for TiNiSn [55,56]. However, $k_l$ has been found to be comparatively low as compared to 13.3 W/mK in ZrNiPb obtained from Phonon3py code [52] and others such as 14.5 W/mK in ZrNiPb, 10-15 W/mK in ZrNiSn [54].

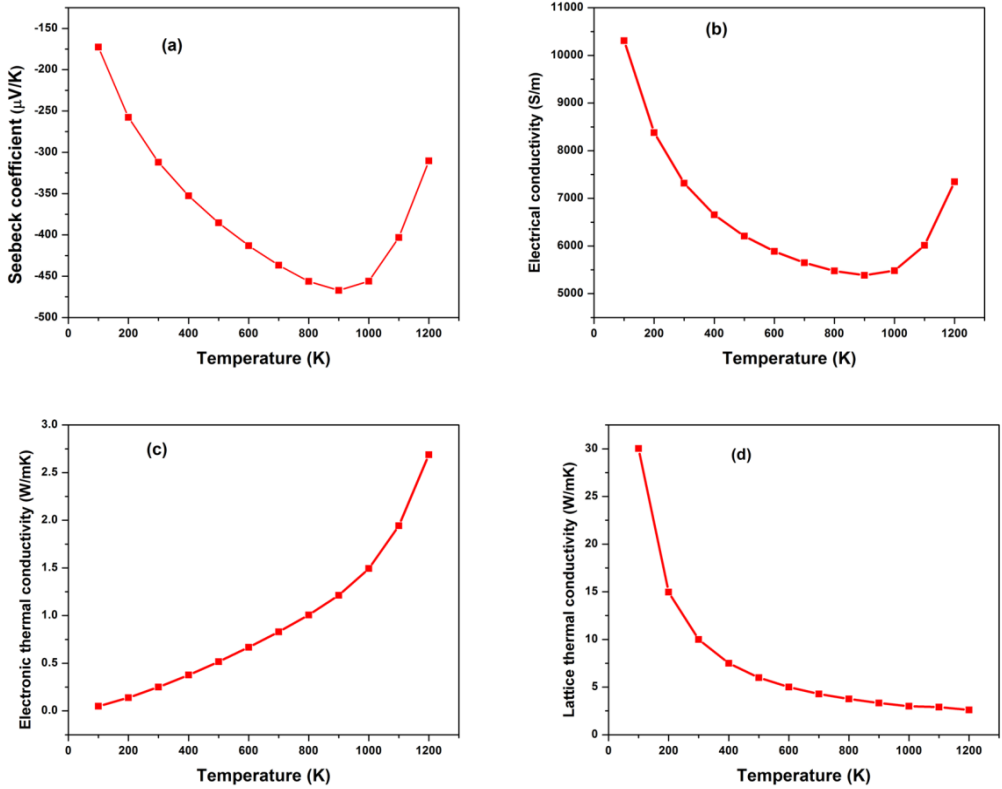

**Fig.4:** Thermoelectric parameters: (a) Seebeck coefficient (b) Electrical conductivity (c) Electron thermal conductivity and (d) lattice thermal conductivity as a function of temperature.

Knowing all the transport parameters such as Seebeck coefficient, electrical conductivity and total thermal conductivity, a dimensionless figure of merit *ZT* has been calculated and presented in Fig. 5. The maximum value of ZT is obtained at 1000 K which is equal to 0.25.

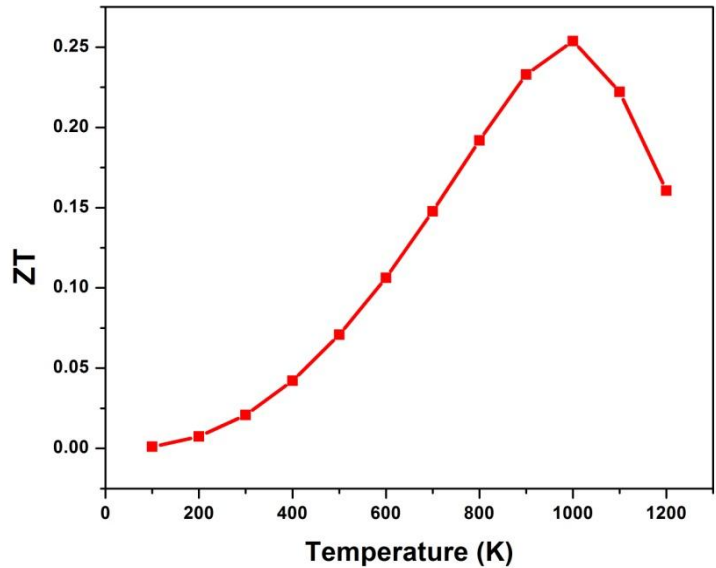

Fig 5: Variation of Figure of merit with temperature

These excellent thermoelectric results indicate that this compound is a good half heusler compound and shall prove to be a very promising material for thermoelectric applications in future.

## Conclusion

The structural, mechanical, electronic, phonon and thermoelectric properties of HfPtPb have been studied for the first time within the frame work of DFT. The calculated equilibrium lattice constant is found to agree well with the previously reported value. The elastic constant values and phonon calculations indicate that, HfPtPb is mechanically and thermodynamically stable. The electronic band structure calculation indicates it to be a direct band gap semiconductor. In relation to energy band, we have computed the thermoelectric properties. A theoretical analysis presented here predicts HfPtPb to be a possible novel thermoelectric material with low value of lattice thermal conductivity, ~ 9.9 W/mK at room temperature. The maximum value of ZT=0.25 has been obtained at 1000 K. We believe that our theoretical results may provide a good reference data to experimentalists to design nanostructure HH compounds exhibiting low thermal conductivity for thermoelectric applications.

## Acknowledgement

D P Rai gratefully acknowledges the financial support received from INSA, New Delhi in the form of INSA visiting fellowship (Ref No. SP/VF-7/2016-17) for this work. Kulwinder Kaur thanks Council of Scientific & Industrial Research (***CSIR***), India for providing fellowship.